\documentstyle[sprocl,epsfig]{article}

\def\vk{{\bf k}}

\def\bea{\begin{eqnarray}}
\def\eea{\end{eqnarray}}
\def\beq{\begin{equation}}
\def\eeq{\end{equation}}
\def\be{\begin{equation}}
\def\ee{\end{equation}}
\def\bc{\begin{center}}
\def\ec{\end{center}}

\def\t0{\tau_0}

\def\bea{\begin{eqnarray}}
\def\eea{\end{eqnarray}}
\def\l({\left(}
\def\r){\right)}

\begin{document}

\title{CORRELATIONS OF MASS-SHIFTED BOSONS}

\author{{\underline{T. CS\"ORG\H O}$^1$} and {M. GYULASSY$^2$} \\[1.5ex]
}

\address{$^1$MTA KFKI RMKI,\\ H-1525  Budapest 114, POB 49, Hungary \\
	$^2$Department of Physics, Columbia University,\\ 
	538 W, 120-th Street, New York, NY 110027, USA\\
E-mails: csorgo@sunserv.kfki.hu, gyulassy@nt1.phys.columbia.edu} 
\maketitle
\abstracts{ We discuss a new kind of correlation, the correlation
of particle-antiparticle pairs emitted back-to-back in the local rest frame
of a medium, if medium effects cause mass modification
of the quanta. The theory of these new correlations 
is formulated for relativistically expanding, locally
thermalized sources.
}

\section{Introduction}
If hadronic
mass-shifts in medium are accompanied with a sudden freeze-out,
they were shown to result in squeezing of the fields~\cite{aw,ac} 
and as a consequence, new kind of back-to-back  correlations 
of the observed quanta may appear~\cite{ac,eger}. 
The theoretical description of these back-to-back correlations
were given first for static sources in refs.~\cite{ac,eger}, 
however, the results were limited  for those
particles that are their own anti-particles, 
like the $\phi$-s or $\pi^0$-s.
The description of back-to-back correlations
was generalized recently to particle - anti-particle pairs
in refs.~\cite{cf98-and,hiro-mina} and in ref.~\cite{acg}.
In the present contribution we discuss  these new kind of 
back-to-back correlations 
following the lines of ref.~\cite{acg}.
We assume the validity of relativistic
hydrodynamics up to  freezeout. 
The local temperature field $T(x)$, the chemical potential distribution
$\mu(x)$,
the four-velocity distribution $u^{\mu}(x)$ 
and the oriented volume elements on freeze-out hypersurface
$d^3\Sigma^{\mu}(x)$ are given 
on the freeze-out hypersurface $\Sigma^{\mu}$, the   
space-time points are denoted by $x = (t, {\bf r})$.
The temperature, flow and chemical potential distributions 
can be taken from realistic hydrodynamical calculations,
for example, from refs.~\cite{rg,cf98-schlei}.

\section{Model Assumptions:} 

The locally thermalized, relativistically expanding
hydrodynamical fluid  ensemble  is specified by the density matrix
\begin{equation}
\hat{\rho} = {1 \over Z} \mbox{\rm e}^{- \frac{1}{( 2 \pi )^3} 
\int\! {d^4k \int d^3 \Sigma^{\mu}(x) k_{\mu}} \, \hat{{\cal W}}
(x,k) \;[k^{\nu}u_\nu(x) - \mu(x)]/T(x) } .
\label{e:hydro}
\end{equation}
 Here $\hat{{\cal W}}$ is the covariant
Wigner operator\cite{degroot}, whose expectation value
is the covariant phase space density, 
$\langle \hat{{\cal W}}(x,p)\rangle 
=2\theta(p^0)\delta(p^2-m^2)f(x,p)$, of on-shell $a$-quanta
in the semi-classical limit.

If we subdivide the fluid into locally thermalized,
macroscopically infinitezimal fluid cells, 
the above hydrodynamic density matrix for locally thermalized systems 
can also be rewritten in the finite form
\be
\rho = \prod_{i} \rho_i,
\ee
where the index $i$ runs over all fluid cells.
The locally thermalized density matrix
in cell $i$ is 
\be
\rho_i = {1 \over Z_i}\exp\left( - (H_i - \mu_i N_i) / T_i \right) .
\ee
Consider, in the rest frame of each of the fluid elements,
the following model Hamiltonian, 
\begin{equation}
{H_i} =  H_{0,i} - \frac{1}{2} \int d^3 {\bf x} d^3 {\bf y} \phi_i({\bf x})
\delta M_i^2({\bf x}-{\bf y}) \phi_i({\bf y}),
\label{ham}
\end{equation}
where $H_{0,i}$ is
the asymptotic Hamiltonian,
\begin{equation}
        H_{0,i} = \frac{1}{2} \int d^3 {\bf x} \left(
                \dot{\phi_i}^2+ |\nabla \phi_i|^2
                +
                        m_0^2 \phi_i^2  \right).
\end{equation}
The scalar field $\phi_i({\bf x})$ in this Hamiltonian, $H_i$, corresponds to 
quasi - particles that propagate in fluid element $i$ with a momentum-dependent
medium-modified effective mass,
which is related to the vacuum  mass, $m_0$,  via
\be
 m_{*i}^2({|{\bf k}|}) =  m_0^2 - \delta M_i^2({|{\bf k}|}).
\ee
The mass-shift is assumed to be limited to long wavelength 
collective modes in every fluid element:
\be
\delta M_i^2({|{\bf k}|}) \ll m_0^2 \qquad \mbox{\rm if} \quad |{\bf k}| > \Lambda_s.
\ee
Here $\Lambda_s$ stands for the scale above which the medium effects
resulting in mass modifications become negligible.

We assume a sudden freeze-out. Mathematically this implies that before the
break-up time $H_i$ governs the time evolution of the fields in each fluid
element, while after freeze-out, the quanta acquire their vacuum mass
$m_0$ and the asymptotic Hamiltonian $H_0$ governs the time evolution.
A more gradual freeze-out can be described by assuming a time-dependent
mass-shift as in ref.~\cite{hiro-mina}. 
In each cell, the post-freeze-out field can be expanded with 
creation and annihilation operators as 
\bea
\phi_i(x) & = & \sum_\vk  (a_{i,\bf k}^{\phantom{\dagger}} \varphi_{i,\vk}(x)
+h.c.),
\eea 
a concept introduced to the field of Bose-Einstein correlation studies
by Si\-nyu\-kov and Makhlin in 1986-1988.

However, before the freeze-out the $a$ quanta do not diagonalize the 
density matrix, due to the mass modification in the medium.
It is possible to introduce new quanta in each fluid cell that
diagonalize the density matrix of the medium
(and {\it not} the Hamiltonian of the medium). These quanta are denoted by
$b_{i, \bf k}$ and they are connected to the $a_{i,\bf k}$ 
quanta by a local two-mode Bogoljubov transformation as given in
ref.~\cite{acg}.

Although the $a$ quanta are observed,
it is the $b$ quanta that are thermalized in medium. 
The observable invariant
single-particle and two-particle momentum distributions are given as:
\begin{eqnarray}
N_1({\bf k}_1) & = & \omega_{{\bf k}_1}{d^3N \over d{\bf k}_1} 
	= \omega_{{\bf k}_1} \langle
 a^\dagger_{{\bf k}_1} a^{\phantom\dagger}_{{\bf k}_1}\rangle , \\
N_2({\bf k}_1,{\bf k}_2) & = & 
\omega_{{\bf k}_1} \omega_{{\bf k}_2} 
        \langle a^\dagger_{{\bf k}_1} a^\dagger_{{\bf k}_2} 
        a^{\phantom\dagger}_{{\bf k}_2}
        a^{\phantom\dagger}_{{\bf k}_1} \rangle ,\\
\hspace{-0.5cm}
\langle a^\dagger_{{\bf k}_1} a^\dagger_{{\bf k}_2} 
a^{\phantom\dagger}_{{\bf k}_2} a^{\phantom\dagger}_{{\bf k}_1} \rangle 
& = &  
\langle a^\dagger_{{\bf k}_1} a^{\phantom\dagger}_{{\bf k}_1}\rangle
\langle  a^\dagger_{{\bf k}_2} a^{\phantom\dagger}_{{\bf k}_2} \rangle 
		\!\! + \!\!
\langle a^\dagger_{{\bf k}_1} a^{\phantom\dagger}_{{\bf k}_2}\rangle
\langle  a^\dagger_{{\bf k}_2} a^{\phantom\dagger}_{{\bf k}_1} \rangle 
		\!\! +  \!\!
\langle a^\dagger_{{\bf k}_1} a^\dagger_{{\bf k}_2}\rangle
\langle  a^{\phantom\dagger}_{{\bf k}_2}
         a^{\phantom\dagger}_{{\bf k}_1} \rangle,
\label{rand}
\end{eqnarray}
where $a_{\bf k}$ is the annihilation operator for the
asymptotic quantum with four-momentum $k^{\mu}\, = \, (\omega_{\bf k},{\bf k})$,
$\omega_{\bf k}=\sqrt{m^2 + {\bf k}^2}$  and
the expectation value of an operator $\hat{O}$ is given by
the medium-modified density matrix $\hat{\rho_b}$ as 
$\langle \hat{O} \rangle = {\rm Tr} \, \hat{\rho_b}\, \hat{O}$.
Eq.(\ref{rand})
has been derived as a generalization
of Wick's theorem for {\em locally} equilibriated (chaotic) 
systems in ref.\cite{sm}. The medium modified density matrix
$\hat{\rho}_b$ has the same functional form as eq.~(\ref{e:hydro}),
but for the medium-modified, mass-shifted $b$ quanta.
In order to simplify later notation, we introduce the 
chaotic and squeezed amplitudes~\cite{acg} defined, respectively, as
\begin{eqnarray}
G_c(1,2) & = & 
\sqrt{\omega_{{\bf k}_1} \omega_{{\bf k}_2} }
  \langle a^{\dagger}_{{\bf k}_1} a^{\phantom\dagger}_{{\bf k}_2}\rangle,\\
G_s(1,2) & = & 
\sqrt{\omega_{{\bf k}_1} \omega_{{\bf k}_2} }
  \langle a^{\phantom\dagger}_{{\bf k}_1}
  a^{\phantom\dagger}_{{\bf k}_2} \rangle .
\label{gs}
\end{eqnarray}
In most situations, the chaotic amplitude, $G_c(1,2) \equiv G(1,2)$
is dominant, and carries the Bose-Einstein correlations,
while the squeezed amplitude, $G_s(1,2)$ vanishes.
In this case,  we recover from (\ref{rand})
the well-known two-particle inclusive correlation function given by
\begin{equation}
C_2({\bf k}_1,{\bf k}_2)  =  {N_2({\bf k}_1,{\bf k}_2)
\over N_1({\bf k}_1) N_1({\bf k}_2) } = 
        1 + { | G(1,2) |^2 \over G(1,1) G(2,2) },
\label{hbt}
\end{equation}
which includes the effect of the two body correlations arising
from the symmetrization of an ideal Bose gas.

For the hydrodynamic ensemble (\ref{e:hydro}), eq.(\ref{hbt}) reduces
to the special form derived by Makhlin and Sinyukov\cite{sm}.
In that case,
the off-diagonal number amplitude reads as 
\begin{equation}
G(1,2) = { \displaystyle\phantom{|}
{1 \over\displaystyle\phantom{|}
(2\pi)^3
}} \,
\int  d^3 \Sigma_\mu K^\mu_{1,2} e^{i q_{1,2}^\nu x_\nu}\; f(x,K_{1,2})
\; \; .\label{num}
\end{equation}
The quantization relation, 
$[a^{\phantom{\dagger}}_{{\bf k}_1},a^\dagger_{{\bf k}_2}] =
\delta^3({\bf q}_{1,2})$, 
gives
\begin{equation}
\langle a^{\phantom{\dagger}}_{{\bf k}_2} a^\dagger_{{\bf k}_1}\rangle
\propto \int  d^3 \Sigma_\mu K^\mu_{1,2} e^{i q_{1,2}^\nu x_\nu} \;
\left[f(x,K_{1,2}) + 1 \right] . \label{numd}
\end{equation}
Note that the relative and average pair
momentum coordinates, ${\bf q}_{1,2}={\bf k}_1-{\bf k}_2,
{\bf K}_{1,2}={\textstyle\frac{1}{2}}({\bf k}_1+{\bf k}_2)$,
and $q^0_{1,2}=\omega_1-\omega_2$
appear in
(\ref{num},\ref{numd}). 
The validity of the approximations leading to
(\ref{num},\ref{numd}) 
requires  the width of $G(1,2)$ as a function
of the relative momentum, $q = |{\bf q}_{1,2}|$, to be small. That width 
is given by $\sim 1/R$, where $R$ is
a characteristic dimension of
the system. The semi-classical limit corresponds
to $KR\gg 1$, where $K$ is $|{\bf K}_{1,2}|$.
Note that $\sqrt{\omega_{k_1}  \omega_{k_2}}\sim K^0_{1,2}$
in this case.
For $qR< 1$, the second term in (\ref{rand})
describes the minimal quantum interference
associated with the indistinguishability of the bosons.
The integration over
the freeze-out three volume surface, $\Sigma^\mu(x)$, of the fluid
is implemented with the invariant measure
$d^3 \Sigma_\mu K^\mu_{1,2}$
that reduces to $d^3\Sigma^\mu K_\mu=K^0 d^3x$ 
in the special case of a constant
freeze-out time.

We can evaluate the chaotic and the squeezed amplitudes 
using the generalization of
(\ref{num},\ref{numd}),
noting that the $b$-quanta satisfy eq.~(\ref{num}) locally in each
fluid cell,
\begin{eqnarray}
G_c(1,2) &=&  {\displaystyle\phantom{|}
        1\over\displaystyle\phantom{|} (2\pi)^3}  
        \int\!\! d^3\Sigma_\mu K^{* \mu}_{1,2}
e^{iq^{* \nu}_{1,2}x_\nu}  |c_{1,2}|^2 n_{1,2} \nonumber \\
        & + &
        {\displaystyle\phantom{|} 1\over\displaystyle \phantom{|}(2\pi)^3}   
        \displaystyle\phantom{|}\int\!\! d^3\Sigma_\mu K^{* \mu}_{-1,-2} 
        e^{iq^{* \nu}_{-1,-2}x_\nu}\; 
        |s_{-1,-2}|^2 \, (n_{-1,-2}+1), \label{e:gc}  \\
G_s(1,2)  \! &=&\!
        {\displaystyle\phantom{|} 1\over\displaystyle\phantom{|} (2\pi)^3} 
         \int\!\! d^3\Sigma_\mu K^{* \mu}_{1,-2}
        e^{iq^{* \nu}_{1,-2}x_{\nu}} \;c_{1,-2}
        s_{1,-2}^*n_{1,-2} 
\nonumber \\
        & + &
         {\displaystyle\phantom{|}
        1 \over\displaystyle\phantom{|} (2\pi)^3} 
        \!\! \displaystyle\phantom{|}\int\! d^3\Sigma_\mu
        K^{* \mu}_{2,-1}e^{iq^{* \nu}_{2,-1}x_\nu}\; 
        c_{2,-1} s^*_{2,-1}\, (n_{2,-1} + 1)  
        . \label{e:gs}
\end{eqnarray}
The above expressions are invariant and defined  even in the
presence of  a non-vanishing flow, as given below:
The local squeezing parameter depends on a generalized energy ratio as 
\begin{eqnarray} 
r(a,b,x) & = & \frac{1}{2}\log\left[{ K^{\mu}_{a,b} u_\mu(x) - \mu(x) 
        \over K^{* \mu}_{a,b}(x) u_\mu(x) - \mu(x)} \right],
        \label{e:rxk}\\
c_{a,b} & =  & \cosh[r({a,b},x)] , \quad 
s_{a,b} \, = \, \sinh[r({a,b},x)],
\end{eqnarray} 
where $a,b = \pm 1,\pm 2$, the mean and the relative momenta
for mass-shifted quanta are defined as  
$K^{* \mu}_{a,b}(x) = [k^{* \mu}_a(x) + k^{* \mu}_b(x) ]/2 $ and
$q^*_{a,b} = k^{* \mu}_a(x) - k^{* \mu}_b(x) . $

In order to define the mass-shifted four-momenta and their back-to-back
four-momenta in a covariant manner, we introduce
\begin{eqnarray}
\tilde{k}^\mu(x) & = & k^\mu - u^\mu(x) [k^\nu u_\nu(x)], \\
\Omega_k(x) & = & u^\mu(x) k^{*}_\mu \, = \, 
	\sqrt{m_0^2 + \tilde{k}^\mu\tilde{k}_\mu -
\delta M^2_x(\tilde{k}^\mu\tilde{k}_\mu) }, \\
k^\mu_-(x) & = & 2 u^\mu(x) [k^\nu u_\nu(x) ] - k^\mu,\\
k^{*,\mu}(x) & = & u^\mu(x) \Omega_k(x) + \tilde{k}^\mu(x),\\
k^{*,\mu}_-(x) & = & u^\mu(x) \Omega_k(x) - \tilde{k}^\mu(x),
\end{eqnarray}
and use the short-hand notation $k^{\mu}_{-1} \equiv k^{\mu}_{1,-}$,
$k^{* \mu}_{-1} \equiv k^{* \mu}_{1,-}$.

\begin{center}
\vspace*{7.5cm}
\includegraphics{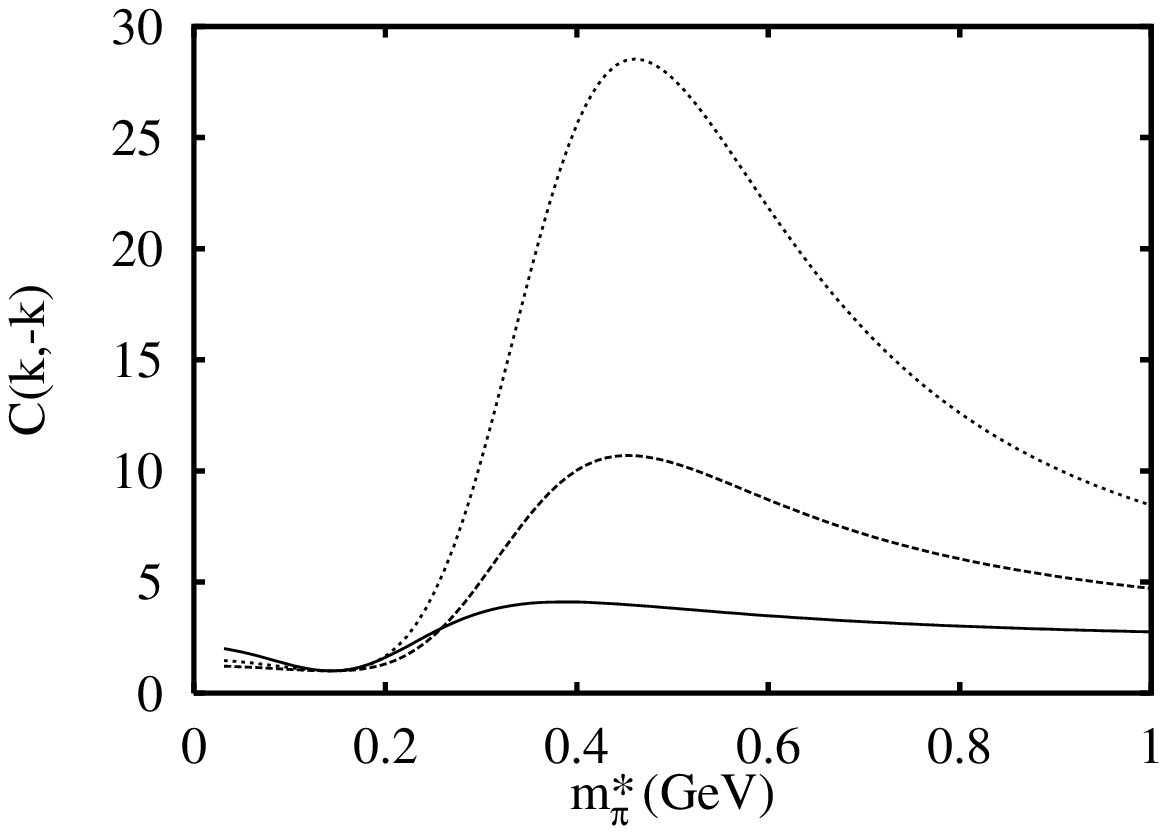}
\begin{minipage}[t]{11.554cm}
{\small {\bf Fig.~1.}  
Dependence of the strength of the back-to-back correlation function
on the medium modified $\pi$ meson mass, $m^*_{\pi}$, for $\pi$ mesons
at $|{\bf k}| = 0 $, 300 and 500 MeV.} This effect can be searched for
both in $(\pi^0,\pi^0)$ and 
in $(\pi^+,\pi^-)$ correlation measurements at BNL or at CERN.
\end{minipage}
\medskip
\end{center}

In eqs. ~(\ref{e:gc},\ref{e:gs}) the Bose-Einstein distribution 
$n_{a,b}$ is 
\begin{equation}
n_{a,b} (x) = {1 \over 
        \exp\left[ {K^{* \mu}_{a,b}(x) u_{\mu}(x) -  \mu(x) \over T(x) }\right] - 1 },
\end{equation}
If the mass-shift is non-vanishing in a finite medium,
as a consequence, a local
squeezing is present ~(\ref{e:rxk}), therefore
the following new expressions are found for the 
particle spectra and the correlation function:
\begin{eqnarray}
        N_1({\bf k}_1) & = & G_c(1,1), \\
        N_2({\bf k}_1,{\bf k}_2) &=& G_c(1,1) G_c(2,2) +
                 |G_c(1,2)|^2 +  |G_s(1,2)|^2, \\
C_2({\bf k}_1,{\bf k}_2) & =  &
        1 
        + 
        {|G_c(1,2) |^2 \over G_c(1,1) G_c(2,2) }
        + {|G_s(1,2)|^2\over G_c(1,1) G_c(2,2) } . 
\end{eqnarray}

\subsection{Particle-antiparticle correlations}
As the Bogoliubov transformation mixes particles
with anti-particles~\cite{cf98-and,weiner-prl,and-ijm}, the above considerations hold only
for particles that are their 
own anti-particles, e.g. the $\phi$ meson and $\pi^0$.
However, the extension to particle -- anti-particle correlations is 
straightforward. 
Let $+$ label particles,
$-$ antiparticles if antiparticle is different from particle,
let $0$ label both particle and antiparticle if they are identical
particles. The non-trivial correlations from mass-modification
for pairs of $(++)$, $(+-)$ and $(00)$ type read as follows:
\begin{eqnarray}
C^{++}_2({\bf k}_1,{\bf k}_2) & =  & 
        1 
        + 
        {|G_c(1,2) |^2 \over G_c(1,1) G_c(2,2) }, \\
C^{+-}_2({\bf k}_1,{\bf k}_2) & =  & 
        1 
        + {|G_s(1,2)|^2\over G_c(1,1) G_c(2,2) } , \\
C^{00}_2({\bf k}_1,{\bf k}_2) & =  &
        1 
        + 
        {|G_c(1,2) |^2 \over G_c(1,1) G_c(2,2) }
        + {|G_s(1,2)|^2\over G_c(1,1) G_c(2,2) } ,  \label{e:cfin} 
\end{eqnarray}
where we have assumed that mass-modifications of particles 
and anti-particles are the same as happens at vanishing baryon density.

 Fig.~1  shows the strengh of the back-to-back correlation function
for mass-shifted $\pi$ mesons. 
The strengh of the back-to-back correlations
is expected to increase with increasing 
momentum of one of the particles from the
back-to-back particle pairs.

The essential properties of the HBT and the back-to-back correlations
are illustrated on Figure~2.
The HBT correlations appear at small relative momenta and the widths of these
correlations measure the lengths of homogeneity, the size of the particle
source within which the momentum space distribution of particles is similar,
a concept introduced by Sinyukov and collaborators ~\cite{sm}.
In contrast, back-to-back correlations measure the lengths of {\it in}homogeneity,
the size of region where particle - anti-particle pairs 
with opposite momenta are created.  
The back-to-back correlations
are limited to a region in momentum space, 
where the  medium mass-modifications
are non-vanishing, as indicated by the cut-off $\Lambda_s$.

\begin{center}
\vspace*{7.5cm}
\null
\includegraphics{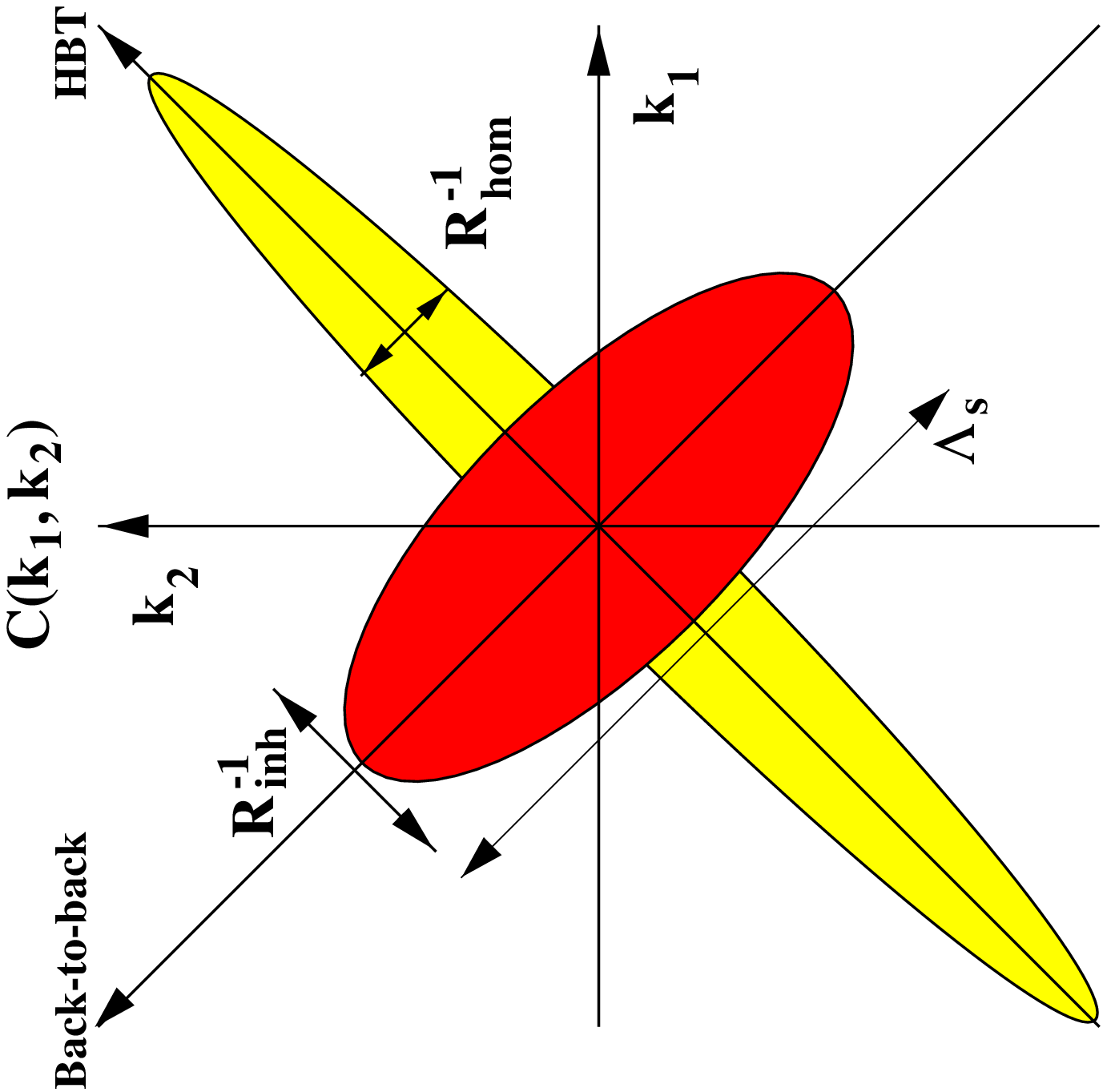}
\begin{minipage}[t]{11.554cm}
\medskip
\medskip
{\small {\bf Fig.~2.}  
Illustation for the essential properties of the back-to-back correlation 
function as created by a modification of hadronic masses in medium.
The back-to-back correlations are present on the scale 
$\Lambda_s > |{\bf k}_1|,|{\bf k}_2|$, and their inverse width
is measuring the lengths of {\it in}homogeneity, $R_{inh}$. }
In contrast, the inverse width of HBT correlations measure the 
lengths of homogeneity $R_{hom}$.
\end{minipage}
\end{center}

\section{Highlights: }
Back-to-back correlations of particle - anti-particle pairs were 
discussed for locally thermalized, medium modified bosons in a covariant
framework. 
These theoretically predicted, new kind of correlations could be looked for
at present and future relativistic heavy ion collisons at CERN  SPS and LHC
as well as at BNL AGS and RHIC experiments. The magnitude of the back-to-back
correlations is expected to be surprisingly large. 
They are expected to carry information about the total longitudinal
extension of inhomogenous, expanding sources.
Our results will have implications also to the signal of DCC 
formation in two-pion correlations~\cite{hiro-mina}:
mass modifications of quanta 
are essential ingredients for a Disoriented Chiral Condensate (DCC) formation
although the thermal averaging performed here is not needed to describe
DCC. Another interesting application of our results could be the
study of particle production via the parametric resonance method 
in the context of reheating in the inflatory Universe~\cite{boyan}.
We have solved the quantum optical problem of squeezing 
in a finite volume as well~\cite{janszky}.

\section*{Acknowledgments}
We thank M. Asakawa for a stimulating and frutiful collaboration
on the back-to-back correlations of mass-shifted bosons. 
We would like to thank I. V. Andreev, V. N. Gribov, S. S. Padula, 
 Yu. M. Sinyukov and R. M. Weiner for stimulating discussions.
This work was supported in part by 
the US - Hungarian  Joint Fund grant MAKA 652/1998, 
by the OTKA grant T026435 and by the NWO  - OTKA grant N25487.

\end{document}